\documentclass[twocolumn,prl,amssymb,epsfig,aps]{revtex4}
\usepackage{epsfig}
\def\ncalcs{230\ }
\def\question{{(Unkn.)\ }}

\begin{document}
\textheight 226.5mm \title{\Large {Uncovering technetium binary ordered structures from first principles}}
\author{
  Ohad Levy$^{1,2}$, Junkai Xue$^1$, Shidong Wang$^1$, Gus L.W. Hart$^{3}$, and
  Stefano Curtarolo$^{1,\star}$}
\affiliation{
  $^1$Department of Mechanical Engineering and Materials Science and
  Department of Physics, Duke University Durham, NC 27708 \\
  $^2$Department of Physics, NRCN, P.O.Box 9001, Beer-Sheva, Israel\\
  $^3$Department of Physics and Astronomy, Brigham Young University, Provo, UT 84602\\
  $^\star${corresponding author, e-mail: stefano@duke.edu} }
\date{\today}
\begin{abstract}
  Technetium, element 43, is the only radioactive transition metal. It occurs naturally on earth in only trace amounts.
  Experimental investigation of its possible compounds is thus inherently difficult and limited.
  Half of the Tc-transition metal systems (14 out of 28) are reported to be phase-separating or lack experimental data.
  Using high-throughput first-principles calculations we present a comprehensive investigation
  of binary alloys of technetium with the transition metals.
  The calculations predict stable ordered structures in nine of these 14 binary systems. They also predict
  unreported compounds in all nine known compound-forming systems and in two of the five systems
  reported with disordered $\chi$ or $\sigma$ phases.
  These results accentuate the incompleteness of the current knowledge on technetium alloys. They
  should guide experimental studies to obtain accurate structural information on potential
  compounds, obviating some of the difficulties associated with such work.

\end{abstract}
\maketitle

Technetium, the only radioactive transition-metal, occurs mainly in spent nuclear fuel.
The practical difficulties of working with a radioactive rare material have so
far hampered studies of technetium solid state physics and chemistry.
In particular, exploration of technetium-based alloys has been very limited.
The available experimental data indicates that 14 of the 28
Tc-transition metal binary systems are non-compound-forming \cite{Pauling,Massalski}. These systems are concentrated in
columns IB, IIB, VIIB and VIIIB of the periodic table. The exceptions in these columns are Zn,
reported to form two compounds with Tc, Mn which is reported with a disordered $\sigma$-phase, and Fe that forms one compound and
the $\sigma$-phase.
Four additional systems, Nb and the column VIB systems, are reported with a disordered $\sigma$-phase.
The seven remaining binary systems of Tc with the transition metals of columns IIIB-VB are reported to be compound-forming.

Recent interest in Tc alloys has been two-fold.  One motivation is a systematic basic research of fourth-row transition-metal alloys
which might provide insight into the existence of general trends in material properties.
Such an example is the recent surprising discovery that
SrTcO$_3$ differs greatly from its strontium metal oxide neighbors, strontium molybdate and ruthenate. It turns out
to be antiferromagnetic with the highest ordering temperature, roughly 1000K, obtained in a material without
a third-row transition metal, whereas SrMoO$_3$ is paramagnetic
with one of the highest conductivities of the metal
oxides and SrRuO$_3$ is a metallic ferromagnet with a transition temperature
around 160K \cite{EERodriguez2011}.
Another motivation for Tc-alloy investigation is the search for potential candidates for long term nuclear waste
disposal in geological repositories by immobilization of Tc-containing residues \cite{Advanced-nuclear-fuel-processing}.
This requires detailed knowledge of synthesis and properties of Tc alloys. Recent studies of Tc-Zr alloys \cite{Poineau2010}
and Tc deposition on gold \cite{MausolfJECS2011} have been carried out with this purpose in mind.

First-principles calculations based on density functional theory (DFT) provide the theoretical means to complement
lacking empirical data, especially in cases were experiments are difficult.
Several results on technetium and technetium alloys have been published using this approach.
The zero temperature equation of state and elastic constants of pure Tc have been calculated in a study
of bcc, fcc and hcp structures of 76 elements \cite{ShangCMS2010}, and the stability of several more complex structures
has been explored in a similar work \cite{SluiterCalphad06}. These studies verify the stability of the hcp structure
for elmental Tc.
A few comparative studies of technetium mono-carbides and borides , TcC, TcB$_2$ and TcB$_4$, and those
of other transition metals, were aimed at revealing the origin of the well-known super-hardness of the corresponding
tungsten structures \cite{YXWangPSS2008,YXWangAPL2007,MWangAPL2008}. It was found that the Tc compounds are also
potential high hardness materials. A similar study compared the properties of transition metal mononitrides \cite{WChenJAC2010}.

In this paper we present a comprehensive investigation of binary alloys of technetium with the transition metals
using high-throughput first-principles calculations.
High-throughput calculations of material properties based on DFT have
acquired an increasing role in recent years as an important tool for
rational material development \cite{monster,ceder:nature_1998,Johann02,Stucke03,curtarolo:prl_2003_datamining,Fischer06}.
They can be used to explore the phase stability landscape of binary
alloys by calculating the formation enthalpies of a large number of
structures, identifying the minima at various component
concentrations.
These calculations can indicate the possible
existence of hitherto unidentified compounds and metastable structures and their characteristics.
A previous study, using this approach, reported on twelve of the Tc-transition
metal binary systems, as part of a larger review of 80 binary alloys aimed at verifying the accuracy of this method \cite{monster}.
The current work covers all the Tc-transition metal alloys using PAW pseudopotentials \cite{paw} and PBE-GGA
exchange-correlation functionals \cite{PBE}, vs. ultrasoft pseudopotentials and
LDA exchange-correlation functionals used in \cite{monster}, and a more extensive structure database.
It uncovers new ordered
structures in a few of the systems discussed in Ref.\ \cite{monster} and in additional ones where experimental data is
scarce and difficult to obtain.

The calculations were performed with the high-throughput framework
{\small AFLOW} \cite{monster,aflow} based on
{\it ab initio} calculations of the energies by the {\small VASP}
software \cite{kresse_vasp}.
The energies were calculated
at zero temperature and pressure, with spin polarization and without
zero-point motion or lattice vibrations. All crystal structures were
fully relaxed (cell volume and shape and the basis atom coordinates
inside the cell). Numerical convergence to about 1 meV/atom was
ensured by a high energy cutoff (30\% higher than the highest energy
cutoff for the pseudo-potentials of the components) and dense 6000
{\bf k}-point Monkhorst-Pack meshes \cite{monkhorst}.

For each system, we calculated the energies of all the reported
crystal structures \cite{Pauling,Massalski} and approximately \ncalcs
additional structures from the {\small AFLOW} database \cite{aflow},
listed in Ref.\ \cite{PRB_hcp_2010}.  This protocol, of searching many
enumerated derivative structures and exhaustively exploring
experimentally reported structures, is expected to give a reasonable
balance between high-throughput speed and scientific accuracy to
determine miscibility, or lack thereof, in Tc alloys (a detailed
discussion on the reliability of the method appeared in
Refs.~\cite{monster,monsterMg}).  However, the existence of additional
unexpected ground-states among unexplored structures can not be ruled out.

\begin{figure}[t]
  \includegraphics[width=0.45\textwidth]{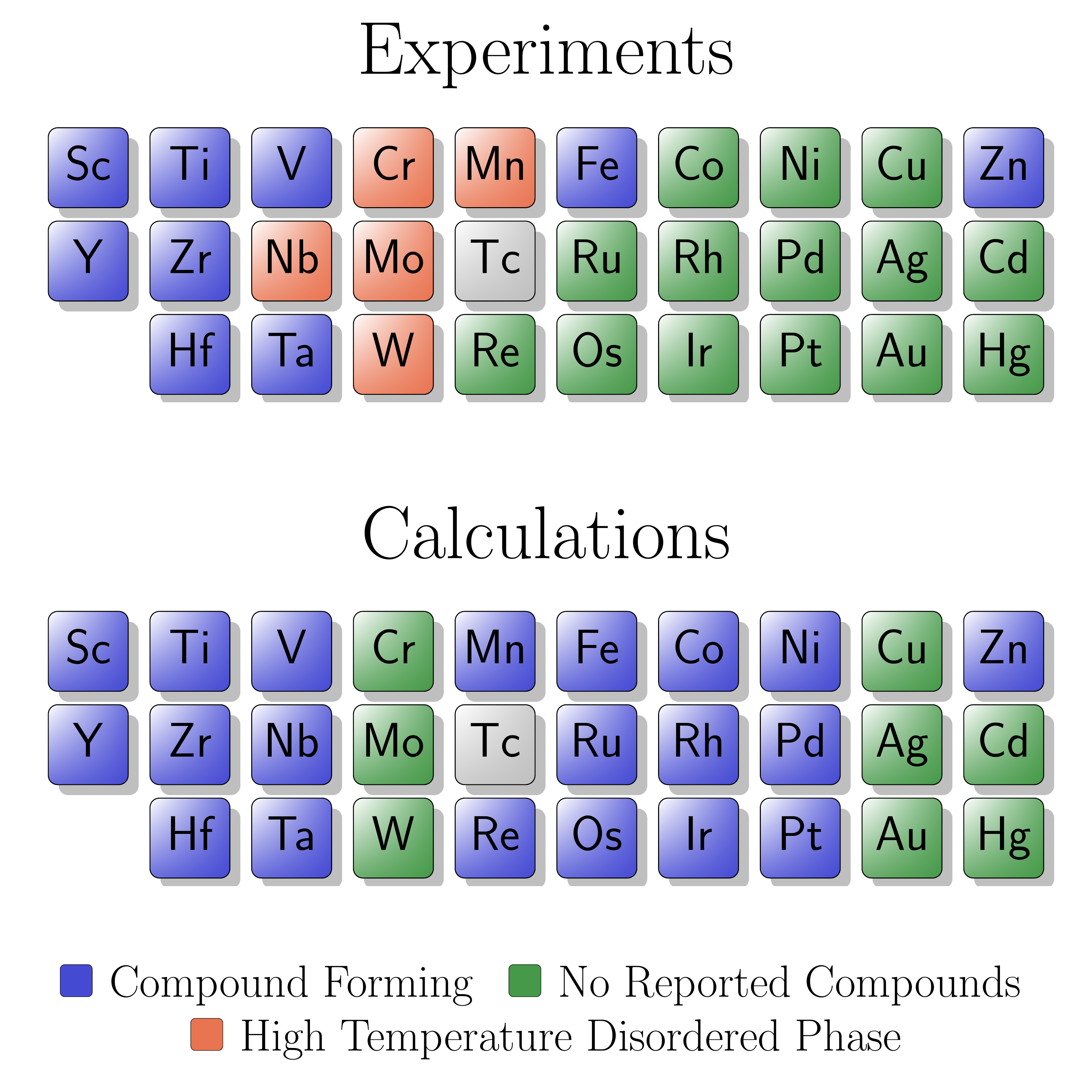}
  \vspace{-3mm}
 \caption{\small
    The phase-separating or compound-forming characteristics of 28 Tc - transition metal alloy systems
    as found in experiments and in {\it ab initio}
    calculations.}
  \label{fig1}
\end{figure}

The calculations reveal stable structures both in systems known to
order and those thought to be phase separating. We show that nine of the 14 technetium
binary intermetallic systems reported as
phase-separating in the experimental literature \cite{Pauling,Massalski} actually
exhibit ordering tendencies, forming stable compounds at
low-temperatures (Fig.~\ref{fig1}). Specifically, we find new structures in four binary systems
not previously studied, Os-Tc, Co-Tc, Ir-Tc and Ni-Tc, and a few
previously unreported stable structures in four systems already
predicted to be ordering in \cite{monster}, Rh-Tc, Ru-Tc, Pt-Tc and Pd-Tc.
The discrepancies between the current predictions and those of Ref.\ \cite{monster},
e.g.\ indication of Tc$_{24}$Ti$_5$ and Tc$_{24}$Zr$_5$ compounds, arise from the larger structure
database scanned in this study.
In addition,
we predict unreported stable structures in all nine binary systems known from experiments to be compound-forming,
and in one of the three systems exhibiting a disordered $\sigma$-phase, Mn-Tc.

%
\begin{table*}[htb]
  \caption{Compounds observed in experiments (``Exper.'') or predicted
    by {\it ab initio} calculations (``Calc.'') in Tc binary alloys (structure prototype in parentheses.
    \question denotes an unknown prototype) and
    their calculated formation enthalpies $\Delta H$. The energy difference between
    a reported structure (unstable in the calculation) and a two-phase tie-line is indicated in
    square parentheses.
    A $\star$ denotes unobserved prototypes described in \cite{monster,PRB_hcp_2010}.
    A $\S$ denotes unobserved prototypes described in Table \ref{table_protos}.
    ``-'' denotes no compounds, and ``N/A" no available data.
    } \label{table1}
  \scriptsize
  \noindent\begin{minipage}{.53\linewidth}\noindent
  \begin{tabular}{c|ccc|c}
    \hline \hline
    & \multicolumn{3}{c|}{Compounds} & $\Delta H$ \\
& Exper.\cite{Pauling,Massalski} & Calc.(Previous)\cite{monster} & Calc.(Present) & meV/at. \\ \hline
 Y  & Tc$_2$Y(C14) & Tc$_2$Y(C14) & Tc$_2$Y(C14) & -176  \\
    &  & TcY$_3$(D0$_{11}$) & TcY$_3$(D0$_{11}$) & -86 \\ \hline
 Sc &  & N/A & Sc$_3$Tc(D0$_{11}$) & -182 \\
    &  &  & Sc$_2$Tc(C11$_b$) & -208 \\
    & ScTc$_2$(C14) &  & ScTc$_2$(C14) & -304 \\
    & Sc$_{0.1}$Tc$_{0.9}$($\chi$) & & Sc$_5$Tc$_{24}$(Re$_{24}$Ti$_5$) & -189  \\ \hline
 Zr & Tc$_{0.88}$Zr$_{0.12}$($\chi$) &  & Tc$_{24}$Zr$_5$(Re$_{24}$Ti$_5$) & -186 \\
    & Tc$_2$Zr(C14) & Tc$_2$Zr(C14) & Tc$_2$Zr(C14) & -314 \\
    & TcZr\question & TcZr(B2) & TcZr(B2) & -356 \\
    &  & TcZr$_2$(C49) & TcZr$_2$(C49) & -271 \\
    &  & TcZr$_4$(D1$_a$) & TcZr$_4$(D1$_a$) & -186 \\ \hline
 Hf &  & N/A & Hf$_3$Tc(Mo$_3$Ti$^\star$)& -269 \\
    &  &  & Hf$_2$Tc(C49) & -357 \\
    & HfTc(B2) & & HfTc(B2) & -482 \\
    & HfTc$_2$(C14) & & HfTc$_2$(C14)& -362 \\
    & Hf$_{0.1}$Tc$_{0.9}$($\chi$) & & Hf$_5$Tc$_{24}$(Re$_{24}$Ti$_5$) & -232 \\ \hline
 Ti & Tc$_{0.9}$Ti$_{0.1}$($\chi$) &  & Tc$_{24}$Ti$_5$(Re$_{24}$Ti$_5$) & -190 \\
    &  & Tc$_2$Ti(C11$_b$) &  & [18] \\
    &  &  & Tc$_5$Ti$_3$(Ga$_3$Pt$_5$) & -416 \\
    & TcTi(B2) & TcTi(B2) & TcTi(B2) & -492 \\
    &  & TcTi$_2$(C49) & TcTi$_2$(C49) & -376 \\
    &  & TcTi$_3$(Mo$_3$Ti$^\star$) & TcTi$_3$(Mo$_3$Ti$^\star$) & -298 \\ \hline
 Nb &  &  & Nb$_5$Tc(HfPd$_5$$^\star$) & -144 \\
    &  & Nb$_3$Tc(Mo$_3$Ti$^\star$) & Nb$_3$Tc(Mo$_3$Ti$^\star$) & -213 \\
    &  & Nb$_2$Tc(C11$_b$) & Nb$_2$Tc(C11$_b$) & -279 \\
    &  & NbTc(B2) & NbTc(B2) & -365 \\
    & Nb$_{0.15}$Tc$_{0.85}$($\chi$) &  & & [19] \\  \hline
 Ta &  & N/A & Ta$_2$Tc(C11$_b$) & -388 \\
    & TaTc(B2) &  & TaTc(B2)  & -501 \\
    & Ta$_{0.15}$Tc$_{0.85}$($\chi$) & & & [36] \\ \hline
 V  & TcV(B2) & N/A & TcV(B2) & -377 \\
    &  & & TcV$_2$(C11$_b$) & -340 \\
    &  & & TcV$_3$(Mo$_3$Ti$^\star$) & -266 \\
    &  & & TcV$_4$(D1$_a$) & -218 \\
 \hline
 \end{tabular}
\end{minipage}
\begin{minipage}{.45\linewidth}\hfill
  \begin{tabular}{c|ccc|c}
    \hline \hline
& \multicolumn{3}{c|}{Compounds} & $\Delta H$ \\
& Exper.\cite{Pauling,Massalski} & Calc.(Previous)\cite{monster} & Calc.(Present) & meV/at. \\ \hline
 Mo & Mo$_{1.5}$Tc$_{2.4}$(A15) & N/A & - &   \\
    & Mo$_{0.3}$Tc$_{0.7}$($\sigma$) & & &   \\
 \hline
 W  & Tc$_{0.7}$W$_{0.3}$($\sigma$) & N/A & - &   \\
\hline
 Cr & Cr$_{0.25}$Tc$_{0.75}$($\sigma$) & N/A & - &   \\  \hline
 Re &  - & N/A & ReTc$_3$(D0$_{11}$) & -5 \\  \hline
 Mn & Mn$_{0.4}$Tc$_{0.6}$($\sigma$) & N/A & Mn$_2$Tc(C14) & -94 \\ \hline
 Fe & FeTc(B2) & N/A & & [158]   \\
    & Fe$_{0.4}$Tc$_{0.6}$($\sigma$) & & FeTc$_{2}$(C16) & -4 \\ \hline
 Os & - & N/A & Os$_3$Tc(D0$_{19}$) & -71 \\
    &  & & OsTc(B19) & -83  \\
    &  & & OsTc$_3$(D0$_{19}$) & -57 \\ \hline
 Ru &  - & Ru$_3$Tc(D0$_{19}$) & Ru$_3$Tc(D0$_{19}$) & -63  \\
    &    & RuTc(B19) & RuTc(B19) & -73 \\
    &    & RuTc$_3$(D0$_{19}$) & RuTc$_3$(D0$_{19}$) & -47 \\
    &    & & RuTc$_5$(RuTc$_5^{\S}$) & -32 \\ \hline
 Co &  -  & N/A & CoTc(B19) & -46 \\
    &    &  & CoTc$_3$(D0$_{19}$) & -53 \\ \hline
 Ir & - & N/A  & Ir$_8$Tc(Pt$_8$Ti) & -89 \\
    &  &   & Ir$_2$Tc(Ir$_2$Tc$^\S$) & -224 \\
    &  &   & IrTc(B19) & -287 \\
    &  &   & IrTc$_3$(D0$_{19}$) & -217  \\  \hline
 Rh &  -  & Rh$_2$Tc(ZrSi$_2$)   & Rh$_2$Tc(Ir$_2$Tc$^\S$) & -157 \\
    &     & RhTc(B19)  & RhTc(B19)  & -175 \\
    &     & RhTc$_3$(D0$_{19}$)  & RhTc$_3$(D0$_{19}$) & -158 \\ \hline
 Ni & - & N/A & Ni$_4$Tc(D1$_a$)  & -30 \\
    &  &  & NiTc$_3$(D0$_{19}$)  & -106 \\ \hline
 Pt & - & Pt$_3$Tc(FCC$_{AB3}^{[001]}$) & Pt$_3$Tc(BCC$_{AB3}^{[001]}$) & -158 \\
    &  &  & Pt$_2$Tc(CuZr$_2$) & -184 \\
    &  & PtTc$_3$(D0$_{19}$) & PtTc$_3$(D0$_{19}$)  & -267 \\ \hline
 Pd & - &  & PdTc(RhRu$^\star$) & -63 \\
    &  & PdTc$_3$(D0$_{19}$) & PdTc$_3$(D0$_{19}$)  & -73 \\ \hline
 Au & - & - & - & \\ \hline
 Ag & - & - & - &  \\  \hline
 Cu & - & - & - & \\ \hline
 Hg & - & - & - & \\ \hline
 Cd & - & - & - & \\  \hline
 Zn &  & N/A  & Tc$_{2}$Zn(FCC$_{AB2}^{[100]}$) & -42 \\
    &  &  & TcZn$_{3}$(L1$_2$) & -62 \\
    & TcZn$_{7}$(CuPt$_7$) &  & TcZn$_{7}$(CuPt$_7$) & -55 \\
    & TcZn$_{15}$\question &  &  & \\
    \hline   \hline
\end{tabular}
\end{minipage}

\end{table*}

\begin{table}[htb]
    \caption{
    Geometry of new prototypes marked by $\S$ in Table \ref{table1}.
    Atomic positions and unit-cell parameters are fully relaxed.}
    \label{table_protos}
    \scriptsize
    {
      \hspace{-8mm}
      \begin{tabular}{||c|c|c||}\hline\hline
        Formula                             &    RuTc$_5$         &      Ir$_2$Tc       \\ \hline
        Lattice                             &    Monoclinic       &      Orthorhombic    \\ \hline
        Space Group (opt.)                  &    Cm No.8 (2)       &      Cmcm No.63      \\ \hline
        Pearson symbol                      &    mS12             &      oS12            \\ \hline
        HT lattice                          &                     & \\
        type/variation \cite{aflowBZ}       & MCLC/MCLC1          &    ORCC/ORCC         \\ \hline
        Conv. Cell                          &                     &                      \\
        $a,b,c$ (\AA)                       & 9.997, 2.752, 6.484 & 2.751,14.374,4.381 \\
        $\alpha,\beta,\gamma$ (deg)         & 90 75.942 90        & 90, 90, 90           \\ \hline
        Wyckoff                             &Ru1 0,0,-0.00140 (2a)&  Ir1 0,0.998,1/4 (4c)\\
        positions                         &Tc1 0.390,0 -0.277 (2a)&  Ir2 0,0.668,1/4 (4c) \\
     \cite{bilbao,tables_crystallography}&Tc2 -0.335,0,-0.331 (2a)& Tc1 0,0.334,1/4 (4c)\\
                                           &Tc3 0.055,0,0.388 (2a)& \\
                                           &Tc4 0.334,0,0.334 (2a)& \\
                                          &Tc5 -0.278,0,0.055 (2a)& \\  \hline
     {\small AFLOW} label \cite{aflow}   & ``128''           &  ``143''         \\  \hline   \hline
      \end{tabular}
    }
\end{table}

The results are summarized in Table \ref{table1}. In the first column,
the 28 alloying metals are ordered according to their Mendeleev number
(or Pettifor's chemical scale) \cite{pettifor:1986}. The
next three columns indicate whether the corresponding binary system is
phase separating or compound forming, according to the experimental
data and to {\it ab initio} calculations reported here and in a previous
study \cite{monster}.

The Pettifor scale is the most successful attempt to date to enable prediction of whether a newly proposed system would
be compound-forming or not and the structure of the expected compounds, based on a single material parameter
\cite{Villarsetal_JAC01}.
Structure maps based on this scale separate well between various reported structures and thus provide a relatively
high degree of predictive insight \cite{pettifor:1986}. However, the maps are purely empirical and their predictive power is
limited by the availability of reliable experimental data (an assessment of the unsatisfactory current situation in this respect is
given in \cite{Villarsetal_JAC01}). It is thus important to complement the sparsity of relevant experimental data
with {\it ab initio} total energy assessments of the competing candidate structures, as we do in this paper.
Ordered by this scale, Table \ref{table1} is divided into three parts with different experimental phase-formation characteristics.
The top (systems Tc-Y to  Tc-V) is occupied almost exclusively by compound-forming systems, except one, Nb-Tc,
which is reported with a disordered $\chi$-phase. The lower part (Os-Tc and below) is almost exclusively occupied
by phase-separating systems, except one compound-forming system at the bottom, Tc-Zn.
The middle part is a border zone of six systems (Mo-Tc to Fe-Tc), four of which exhibit a disordered $\sigma$-phase,
one, Fe-Tc, reported with a single compound in addition to the $\sigma$-phase , and one, Re-Tc, phase-separating.

The picture emerging from the calculations is considerably different. Ordered structures are predicted in the three lower
systems of the of the middle part, Re-Tc, Mn-Tc and Fe-Tc, and in the eight upper systems of the lower part, Os-Tc to Pd-Tc,
thus predicting a cluster of eleven compound-forming systems in the middle of the table.
Stable structures $M$Tc$_3$ of prototype D0$_{19}$ are found for eight
of the nine column VIII transition metals, except Fe. For two of them,
Os and Ru, the structure $M$$_3$Tc with the same prototype is also stable.
Three of the compounds in this cluster, RuTc$_5$, Ir$_2$Tc and Rh$_2$Tc, are predicted with crystal
structures that have no known prototype or {\it Strukturbericht}
designation. They were found among the symmetrically
distinct fcc-, bcc- and hcp-based superstructures included in the {\small AFLOW} database \cite{monster}. 
These sructures are described in Table \ref{table_protos}.

We also find stable ordered structures in Nb-Tc, indicating
a continuous cluster of eight compound-forming systems at the top. Within this cluster,
the metals of the VB column order into structure $M$$_2$Tc of prototype C11$_{b}$, and those of the IIIB and IVB columns,
except Ti, form a stable structure $M$Tc$_2$ of prototype C14.
Six of these eight systems are reported with the disordered
$\chi$-phase ({\it Strukturbericht} A12, space group I$\bar{4}3$m). In four of them we find stable structures
of prototype Re$_{24}$Ti$_5$, which is an ordered realization of this phase. This indicates that the $\chi$-phase regime of stability
extends to the low-temperature region of these binary phase-diagrams. In the other two systems, Nb-Tc and Ta-Tc, this structure has
a higher formation enthalpy than the two-phase region tie-line, indicating decomposition of the $\chi$-phase at low-temperatures.
Experimental studies of the Tc-Zr system report a
structure, denoted Tc$_6$Zr, with the crystalographic characteristics
of the $\chi$-phase and a wide range of stoichiometries \cite{Poineau2010}. Our calculations thus identify the prototype and confirm
the existence of a corresponding ordered structure at low temperatures. Similar behavior should be expected
in the adjacent systems, Sc-Tc, Hf-Tc and Tc-Ti.

The remaining phase-separating systems form two small groups. One in the middle of the table, of three systems reported with
a $\sigma$-phase and predicted to have no ordered stable structures. The other,  near the bottom, includes five systems
for which both experiments and calculations indicate no compound formation.
The Tc-Zn system remains an isolated compound-forming system at the
last row of Table \ref{table1}.   For this system, the calculations
predict two new ordered structures,  Tc$_{2}$Zn and TcZn$_{3}$  in
addition to the observed one TcZn$_{7}$. We find no stable structures in the vicinity of
TcZn$_{15}$, reported in experiments with an unidentified prototype.

The almost perfect grouping of systems into four well-defined clusters by their predicted
phase-formation characteristics nicely complements the
trends indicated by the Pettifor chemical scale. It reverses the ratio of phase-separating
to compound-forming systems from the experimental database.

Empirical data on technetium alloy properties is incomplete and difficult to obtain due
the radioactivity of the element.  Generating such data using {\it ab initio} electronic structure
calculations is thus of special interest.
In this paper, we present results of a computational high-throughput
study on phase ordering in Tc alloys that are considerably
different from those reported in current experimental data. These
theoretical predictions should serve as guide for future studies of these materials and as
the starting point for designing desirable alloys for various potential applications.


Research supported by ONR (N00014-11-1-0136, N00014-09-1-0921), and NSF (DMR-0639822, DMR-0650406).  We are
grateful for extensive use of the Fulton Supercomputer Center at Brigham Young University and Teragrid resources (MCA-07S005).

\end{document}